\documentclass[aps,prd,twocolumn]{revtex4}
\usepackage{amsmath,bm,graphicx,thmsa}
\begin{document}

\title{Estimates for the characteristic problem of the first-order
reduction of the wave equation}

\author{Simonetta Frittelli}
\email[]{simo@mayu.physics.duq.edu}
\affiliation{Department of Physics, Duquesne University,
        Pittsburgh, PA 15282}

\date{\today}

\begin{abstract}

We calculate certain estimates for the solution of the characteristic
problem of the wave equation reduced to first order, in terms of the
free data prescribed on two transverse surfaces, one of which is
characteristic. Estimates of such kind ensure the stability of the
solutions under small variations of the data. Similar estimates exist
for the derivatives of the solution as well.

\end{abstract}
\maketitle

\section{Well posedness and characteristic problems} \label{sec:1}

Given a system of partial differential equations where a unique
solution exists for some given data, oftentimes it is necessary to
know how sensitive the solution is to small variations of the
data. One would not want the equations to amplify any uncertainty
in the data beyond control. When, in addition to existence and
uniqueness, the equations guarantee the stability of the solution
under small perturbations of the data, the problem is referred to
as \textit{well posed}~\cite{courant}.

There are two components to a well-posed problem. First there are
the equations themselves, which are usually defined only up to
transformations of variables and coordinates -- respecting the
differential order of the equations. Second, there is the data
set, usually an appropriate set of values prescribed on a chosen
surface. In coordinate-independent terms, the choice of data
surface determines the type of problem for the given equations. If
one seeks a solution in the interior of a region bounded by a
closed data surface, we have a \textit{boundary problem}. If the
solution is needed on half of the available space of independent
variables ($x^1\ge 0$ in the appropriate choice of coordinates for
the problem), then we have an \textit{initial value problem}, or
``Cauchy problem''. It should come as no surprise that some
equations have at the same time a well-posed problem and an
ill-posed problem. The textbook example is the two-dimensional
Laplace equation $u_{xx}+u_{yy}=0$, for which the boundary problem
is well posed but the initial value problem is not. A less
acknowledged but equally powerful example is the wave equation
$u_{xx}-u_{tt}=0$, for which the reverse statement holds.
Typically elliptic equations admit well-posed boundary problems,
whereas hyperbolic equations admit well-posed initial value
problems.

In the case of hyperbolic equations, there are also
\textit{characteristic problems}. In this case, for a unique
solution to exist, the data must be prescribed on two intersecting
transverse surfaces, one of which is ruled by characteristic
curves and we assume the other one is not~\cite{duff}. On the
side, characteristic problems are not to be confused with mixed
initial-boundary value problems, where values are prescribed on
two intersecting surfaces but such that the initial surface is not
characteristic, because in such mixed problems the number of data
variables necessary to specify a unique solution is larger than in
the initial value problem of the same equations. The
characteristic problem of a set of equations requires the same
number of data as the corresponding initial value problem. In
characteristic problems, the method of solution is an adaptation
by Duff of the Cauchy-Kowalewsky theorem for initial value
problems. As for the stability of the solutions, there does not
seem to have been a great deal of interest. Characteristic
problems are usually neglected in favor of their very powerful
cousins, the initial value problems, which require only one
correctly chosen data surface. Still, in some physical contexts
such as general relativity, characteristic problems have
traditionally been very fruitful in connection with issues of
radiation~\cite{sachs}, and have been used methodically to obtain
numerical solutions~\cite{jefflr}. The question of well-posedness
in such cases becomes quite relevant.

Here the stability of the solutions of the characteristic problem of
the wave equation in three spatial dimensions, reduced to first order,
is explored by means of a particular method.  The method is analogous
to the energy estimates used in the case of initial value problems of
first-order hyperbolic systems~\cite{courant,kreissbook}. The
principle behind the method is to obtain, as a consequence of the
equations, an inequality by which the ``size'' of the solution is
bounded by the ``size'' of the data, namely: to estimate the solution
in terms of the free data. In linear problems, the estimate applied to
the difference between two solutions that are initially close
trivially ensures stability.

In Section~\ref{sec:2} we include a brief expository review of the
method for the initial value problem of the wave equation reduced to
first order. This is done with the purpose of developing a parallel
with either characteristic planes or cones as data surfaces in
Sections~\ref{sec:3} and \ref{sec:4}. It is found that a certain kind
of estimate for the solution in terms of the data can be established
in both cases, demonstrating that these two characteristic problems
for the wave equation are well posed. Section~\ref{sec:5} contains
concluding remarks including a brief overview of the status of the
question of stability of characteristic problems in a more general
sense. We hope to uphold that even classic problems such as the wave
equation have the potential for new insights when viewed from a
different angle.

\section{The Cauchy problem of the
wave equation reduced to first order}
\label{sec:2}

We have the wave equation in cartesian coordinates in three spatial
dimensions given by
\begin{equation}\label{waveeq}
    \psi_{tt} = \psi_{xx} +\psi_{yy} +\psi_{zz}
\end{equation}

\noindent where a subscript $x^a\equiv (t,x,y,z)$ denotes partial
differentiation with respect to the coordinate $x^a$, as usual. We do
not deal with this equation directly, but prefer to ``reduce''
(actually, \textit{extend}) the equation to a system of first order
equations. In doing so, one substitutes the original problem with a
problem that has more solutions, among which the solutions to the
original problem can be singled out.

We define new variables $U\equiv \psi_t, P\equiv \psi_x, Q\equiv
\psi_y$ and $R\equiv \psi_z$. In terms of these variables,
(\ref{waveeq}) is equivalent to the following system
\begin{subequations}\label{udotcart}
\begin{eqnarray}
    U_t    &=&  P_x +Q_y +R_z,   \label{Udotcart}\\
    P_t    &=& U_x,         \label{Pdotcart}\\
    Q_t    &=& U_y,         \label{Qdotcart}\\
    R_t    &=& U_z.         \label{Rdotcart}\\
    \psi_t &=& U,           \label{psidotcart}
\end{eqnarray}
\end{subequations}

\noindent in the sense that all the solutions $\psi(x^a)$ of
(\ref{waveeq}) are solutions of (\ref{udotcart}). In order to single
them out we need to impose the following constraints on the initial
data surface
\begin{subequations}
\begin{eqnarray}
    {\cal C}^x\equiv P-\psi_x&=&0,  \label{constx}\\
    {\cal C}^y\equiv Q-\psi_y&=&0,  \label{consty}\\
    {\cal C}^z\equiv R-\psi_z&=&0,
\end{eqnarray}
\end{subequations}

\noindent which remain satisfied because their $t-$derivatives vanish
(${\cal C}^x{}_t= {\cal C}^y{}_t=  {\cal C}^z{}_t= 0$) by virtue of
(\ref{udotcart}). However, we do not have to restrict the data for the
estimates that follow, so we ignore the constraints from now on.
In matrix notation the system (\ref{udotcart}) has the form
\begin{equation}\label{prepre}
    \bm{A^a} \partial_a v +\bm{D}v =0
\end{equation}

\noindent where $v=(U,P,Q,R,\psi)$,  $\bm{A^t}$ is the identity
matrix, the other three matrices  $\bm{A^i}$ are symmetric,
$\bm{D}$ has all vanishing coefficients but one (of value $-1$),
and summation over repeated indices is understood.  Multiplying on
the left with $v$ we have
\begin{equation}
    v\bm{A^a}\partial_a v +v\bm{D}v =0,
\end{equation}

\noindent and, because the matrices are symmetric, we can extract the
partial derivatives to obtain a ``conservation law'':
\begin{equation}\label{consvlawcart}
    \partial_a(v\bm{A^a}v)+2v\bm{D}v =0.
\end{equation}

\noindent Because we are not interested in a boundary-value problem,
we now assume that the fields either decay sufficiently fast at
infinity so that their volume integration in the entire space $R^3$
is finite, or else they are periodic, so that we can restrict
attention to a box in $R^3$. Equivalently, we assume that the fields
have a Fourier transform or series.  With this assumption we can
first split off the time part of the conservation law
(\ref{consvlawcart}), and then integrate it in the
spatial volume ${\cal V}$ (denoting either $R^3$ or a box):
\begin{equation}\label{preestcart}
    \frac{d}{dt}\int_{\cal V}v\bm{A^t}v \;d^3\!x
    +\int_{\cal V}\partial_i(v\bm{A^i}v) \;d^3\!x
    +2\int_{\cal V}v\bm{D}v \;d^3\!x=0.
\end{equation}

\noindent With our assumptions, the second term evaluates to zero.
In the first term we have the $L_2$ norm of the fields, because
$\bm{A^t}$ is the identity matrix and thus
\begin{equation}
    \int_{\cal V}v\bm{A^t}v \;d^3\!x
    = ||v(t,\cdot)||^2
\end{equation}

\noindent Since $v\bm{D}v =-U\psi$, Eq.~(\ref{preestcart}) is
equivalent to
\begin{equation}
    \frac{d}{dt}||v(t,\cdot)||^2
    =\int_{\cal V}2U\psi \;d^3\!x.
\end{equation}

\noindent Finally, since $2U\psi\le U^2+\psi^2$ then $\int_{\cal
V}2U\psi \;d^3\!x\le ||v(t,\cdot)||$, so we have
\begin{equation}
    \frac{d}{dt}||v(t,\cdot)||^2
    \le ||v(t,\cdot)||^2
\end{equation}

\noindent which implies
\begin{equation}
    ||v(t,\cdot)||^2
    \le e^t||v(0,\cdot)||^2
\end{equation}

\noindent and we have our estimate of the ``size'' of the solution
in terms of the ``size'' of the initial data. Since the system is
linear, the difference of two solutions is also a solution, so
small variations of the initial data result in small variations of
the solution, at least for small enough times. The exponential
factor is there because of the presence of undifferentiated terms
in (\ref{udotcart}). In spite of the exponential factor, the
inequality shows that the solution depends continuously on the
data, so that the variation in the solution at any time can be
controlled by refining the accuracy of the data. Nonetheless, as
usual~\cite{kreissbook}, the inequality is useless for large
values of $t$ for numerical purposes, where the interest resides
in explicitly estimating the error in a given solution that starts
with no less than round-off error.

Alternatively, one can get a better estimate in this case by
treating $\psi$ separately from the rest of the variables. Clearly
the equations for the variables $U,P,Q$ and $R$ decouple from
$\psi$, and have no undifferentiated terms. A solution can be
found independently of the value of $\psi$, and can be estimated
also independently of $\psi$.  If we reproduce the preceding
reasoning with $v\equiv (U,P,Q,R)$, then we have
Eq.~(\ref{prepre}) with $\bm{D}=\bm 0$, which leads to
\begin{equation}
\frac{d}{dt}||v(t,\cdot)||^2 =0,
\end{equation}

\noindent and consequently
\begin{equation}\label{cauchyestimate1}
    ||v(t,\cdot)||^2
    = ||v(0,\cdot)||^2,
\end{equation}

\noindent without an exponential factor. Then $\psi$ can be
estimated by
\begin{equation}
    \psi = \psi|_{t=0} +\int_0^t U dt'
\end{equation}

\noindent which is the solution of $\psi_t=U$ for known source $U$
and given initial data at $z=0$. This leads, for instance, to an
estimate of the form
\begin{equation}
    \psi^2 \le 2\psi|^2_{t=0}
        +2\left(\int_0^t U dt'\right)^2.
\end{equation}

Most remarkably, one can also control the smoothness of the
solution. The spatial derivatives of the fields have estimates in
terms of the spatial derivatives of the initial data. One can see
this easily by taking a spatial derivative $\partial_i$ of the
system (\ref{Udotcart}-\ref{Rdotcart}), which thus becomes an
evolution system for the variables $v_i = (U_i,P_i,Q_i,R_i)$ of
exactly the the same kind as that for $v$, leading to estimates for
each spatial derivative $v_i$ in terms of the same spatial
derivative of the initial data:
\begin{equation}
    ||v_i(t,\cdot)||^2
    = ||v_i(0,\cdot)||^2,
\end{equation}

\noindent for $i=x,y,z$ alternatively.

As a hyperbolic equation, the wave equation admits characteristic
lines at $45^\circ$ with the time axis in all spatial directions
from any fixed point, that is: all lightlike straight lines through
any point. A (hyper)surface that has a null (lightlike) normal at
all points is automatically ruled by characteristic lines and is
referred to as a null surface. A null surface is simply the
evolution of a wavefront forward in time out of any given initial
shape. Any null surface can be chosen as the data surface of a
characteristic problem. Thus there are infinite ways to set up a
characteristic problem for the wave equation.  But there are two
aesthetically appealing cases.  The first case is that of a null
plane acting as the data surface (the evolution of a plane
wavefront). Of course, there are null planes in all possible
directions, so this may not be useful unless our problem has a
particular preferred direction. If there is no particular preferred
direction then null cones are suitable (the evolution of a spherical
wavefront out of some point of origin). These two choices of
characteristic data surfaces lead to similar characteristic problems,
which are examined in the following two Sections.

\section{Well-posedness of the characteristic problem of the wave
equation with a plane characteristic data surface}
\label{sec:3}

The simplest characteristic problem of the wave equation prescribes
data on a plane characteristic surface. In addition, we use a plane
timelike surface to prescribe the necessary complementary data.  In
Subsection~\ref{subsec:3.1} we examine the stability of the solutions
under small perturbations of the data. Estimates for the derivatives
of the solutions are derived in Subsection~\ref{subsec:3.2}.

\subsection{Estimate of the solution}
\label{subsec:3.1}
We transform coordinates $(t,x,y,z)\to (u,x,y,z)$ with
\begin{equation}
    u=t-z,
\end{equation}

\noindent so that the level surfaces of $u$ are null planes. The wave
equation (\ref{waveeq}) in these coordinates reads
\begin{equation}\label{nullcartoriginal}
    2\psi_{uz}
    -\psi_{xx}
    -\psi_{yy}-\psi_{zz} =0.
\end{equation}

\noindent The second derivative with respect to $u$ does not appear
in the equation, so we only need to define three additional variables
to reduce the equation to a first-order system. We have, as before,
$ P\equiv \psi_x, Q\equiv
\psi_y$ and $R\equiv \psi_z$. In terms of these variables,
(\ref{waveeq}) is equivalent to the following system
\begin{subequations}\label{planechar}
\begin{eqnarray}
    2R_u   &=& P_x+Q_y+R_z  ,   \label{Ru}\\
    P_z    &=& R_x      ,   \label{Pz}\\
    Q_z    &=& R_y      ,   \label{Qz}\\
    \psi_z &=& R        ,   \label{psiz}
\end{eqnarray}
\end{subequations}

\noindent in the sense that all the solutions $\psi$ are contained
in the set of solutions of this system.  This system has a unique
solution in the ``wedge'' space with $u\ge 0$ and  $z\ge 0$ if the
value of $R$ is prescribed on the null surface $u=0$ and the values
of $P, Q$ and $\psi$ are prescribed on the surface $z=0$.  This is
referred to as the canonical form of the characteristic problem of the
wave equation~\cite{duff}.  Like all characteristic problems in
canonical form, it consists of two sets of equations: one of them
internal to each characteristic surface
[Eqs.~(\ref{Pz}-\ref{psiz})], and the other moving from one
characteristic surface to the next [Eq.~(\ref{Ru})]. The field $R$
is a \textit{normal variable\/}, whereas $P,Q$ and $\psi$ are
referred to as \textit{null variables\/}~\cite{duff}. We can single
out solutions to the original second-order wave equation by choosing
data so that the constraints (\ref{constx}) and (\ref{consty}) are
satisfied along the surface $z=0$. They remain satisfied everywhere
because $\partial_z{\cal C}^x=\partial_z{\cal C}^y = 0$ by virtue of
(\ref{planechar}).

As in the case of the initial value problem in the preceding
Section, we see that the equations for $(R,P,Q)$ decouple from
$\psi$. Therefore we can find and estimate a solution independently
of $\psi$. Disregarding Eq.~(\ref{psiz}), the rest of the system
(\ref{planechar})  has the form
\begin{equation}
    \bm{A^u} v,_u + \bm{A^z} v,_z+
    \bm{A^x} v,_x+
    \bm{A^y} v,_y=0
\end{equation}
where $v=(R,P,Q)$, and the matrices are given by
\begin{equation}\label{cartnullmatricesuz}
\bm{A^u}
=
\left(\begin{array}{rrr}
    2&0&0\\
    0&0&0\\
    0&0&0\\
      \end{array}
\right)
\hspace{1cm}
\bm{A^z}
=
\left(\begin{array}{rrr}
    -1&0&0\\
    0 &1&0\\
    0 &0&1
      \end{array}
\right)
\end{equation}
\begin{equation}\label{cartnullmatricesxy}
\bm{A^x}
=
\left(\begin{array}{rrr}
    0&0&-1\\
    0&0&0\\
    \!\!\!-1&0&0\\
      \end{array}
\right)
\hspace{0.5cm}
\bm{A^y}
=
\left(\begin{array}{rrr}
    0&0&-1\\
    0&0&0\\
    \!\!-1&0&0
      \end{array}
\right)
\end{equation}

\begin{figure}
 \includegraphics[width=2.3in]{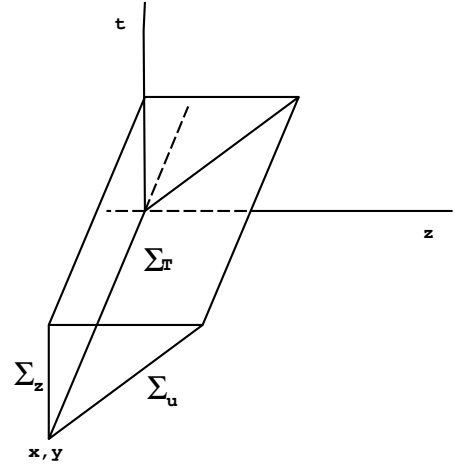}

\caption{The ``hyper-prism'' of integration for  the
characteristic problem with a null plane data surface. The top
boundary surface $\Sigma_T$ lies at $u+z=T$ constant and has
measure $d\Sigma_T = dxdydz/\sqrt{2}$. The surface $\Sigma_u$ is a
null plane at $u=0$ and has measure $d\Sigma_u = dxdydz$. The
surface $\Sigma_z$ lies at $z=0$ and has measure $d\Sigma_z =
dudxdy$. Both surfaces $\Sigma_u$ and $\Sigma_z$ are data surfaces
for the characteristic problem. \label{fig1}}

\end{figure}

\noindent Because all the matrices $\bm{A^a}$ ($a=u,x,y,z$) are
symmetric, it follows that
\begin{equation}\label{nullplanecons}
    \partial_a(v \bm{A^a}v)=0,
\end{equation}

\noindent so we obtain a ``conservation law'' of the form
$\partial_aJ^a=0$. Integration of this conservation law on any
given volume in $R^4$ (of measure $d^4x = dudzdxdy$) gives
us relationships between the flows of $v \bm{A^a}v$ across the
surfaces that enclose such volume. We choose our four-volume of
integration to be a ``hyper-prism'', bounded by the surfaces
$u=0$ (of outward pointing normal  $n^1_a=-\delta^u_a$), $z=0$
(of outward pointing normal  $n^2_a=-\delta^z_a$) and $u+z=T$
for a fixed constant $T$ (of outward pointing normal
$n^1_a=(\delta^u_a+\delta^z_a)/\sqrt{2}$). Assuming that the
fields admit Fourier transforms in the $x$ and $y$ directions,
we let the ``hyper-prism'' extend to infinity along $x$ and $y$,
or alternatively we can have a periodic boundaries in $x$ and
$y$. Our four-volume of integration ${\cal V}^4$ is represented
in Fig.~\ref{fig1}. Integrating on this ``hyper-prism'' we find
\begin{eqnarray}\label{breakup}
0=\int_{{\cal V}^4}
\partial_a(v \bm{A^a}v)d^4\!\!x
&=& \int_{u+z=T}\hspace{-0.8cm}
     v(\bm{A^u}\!+\!\bm{A^z})v \;dzdxdy \nonumber\\
&&-\int_{u=0} \hspace{-0.4cm}v\bm{A^u}v \;dzdxdy\nonumber\\
&&-\int_{z=0} \hspace{-0.4cm}v \bm{A^z}v \;dudxdy
\end{eqnarray}

\noindent Since, by (\ref{cartnullmatricesuz}),
$\bm{A^u}\!+\!\bm{A^z}$ is the identity matrix, then
$v(\bm{A^u}\!+\!\bm{A^z})v = R^2+P^2+Q^2$ and its
integral on the surface $\Sigma_T$ may \textit{define} a
(positive definite) norm of the solution,
\begin{equation}
||v||_T^2 \equiv \int_{u+z=T}\hspace{-0.8cm}
    v(\bm{A^u}\!+\!\bm{A^z})v\;dzdxdy ,
\end{equation}

\noindent which may be used as a measure of the ``size'' of the
solution generated by data on $u=0$ and on $z=0$. On the  other
hand, $v\bm{A^u}v = 2R^2$ and $v \bm{A^z}v=
-R^2+P^2+Q^2$. With these
expressions, (\ref{breakup}) can be written as
\begin{equation}
 ||v||_T^2 = 2\!\int_{\Sigma_u}\hspace{-0.4cm}R^2 d\Sigma_u
       +\int_{\Sigma_z}\hspace{-0.35cm}(-R^2+P^2+Q^2)
        d\Sigma_z
\end{equation}

\noindent which trivially implies the following inequality
\begin{equation}\label{estimatenullplane}
 ||v||_T^2 \le 2\left(\!\int_{\Sigma_u}\hspace{-0.4cm}R^2 d\Sigma_u
       +\int_{\Sigma_z}\hspace{-0.35cm}
        \left(P^2+Q^2\right)
    d\Sigma_z\right).
\end{equation}

\noindent The right-hand side is a positive-definite measure of
the ``size'' of the free data, properly taking into account both
data surfaces. This differs markedly from other possible
energy-like norms that could be defined, mirroring the Cauchy
problem, by integrating the squares of \textit{all fields\/} on
the initial characteristic surface. When all fields are integrated
on the surface at a fixed value of $u$, one is, in a sense,
over-counting the null data, thus a literal translation of the
energy of the Cauchy problem does not, in fact, play the intended
role. On the contrary, the inequality (\ref{estimatenullplane})
functions as the proper characteristic analog of the standard
estimate for initial value problems. Small data $R$ on $u=0$ and
$P,Q$ on $z=0$ result in small values of $R,P$ and $Q$ within the
``wedge'' space $u\ge 0,z\ge 0$.

Once $R$ is known,
we can estimate $\psi$ by integrating Eq.~(\ref{psiz}) with
given values on $z=0$
\begin{equation}
    \psi = \psi|_{z=0} +\int_0^z R dz',
\end{equation}

\noindent which implies, for instance,
\begin{equation}
    \psi^2 \le 2\psi|^2_{z=0}
    +2\left(\int_0^z R dz'\right)^2.
\end{equation}

\subsection{Estimates of the derivatives}
\label{subsec:3.2}

It is not as simple to estimate the derivatives of the variables
in terms of the derivatives of the data as it is in the case of
the Cauchy problem discussed in the previous Section.  However,
with some ingenuity, we can show that the derivatives
$R_x,R_y,R_z,P_x,P_y,P_u,Q_x,Q_y$ and $Q_u$ can be controlled with
the choice of data. These play a role that is analogous to that of
the space derivatives in the case of the Cauchy problem discussed
in Section~\ref{sec:2} -- $P_z,Q_z$ and $R_u$ are already given by
the system of equations (\ref{planechar}), and can be estimated
directly by the use of the system of equations, as in the case of
the Cauchy problem -- . Of these, $R_x,R_y,R_z$ should be treated
as normal variables and the rest as null variables. If so, the
data for $R_x,R_y,R_z$ can be obtained from the derivatives of the
data for $R$ and we avoid the introduction of more arbitrary data
(likewise with the rest of the first-order variables). With this
in mind, we obtain equations for $R_x,R_y,R_z$ by applying
$\partial_i$ to (\ref{Ru}). Every appearance of $P_z$ or $Q_z$ in
the result is substituted in terms of the derivatives of interest
by the use of (\ref{Pz}-\ref{Qz}). For $P_x,Q_x,P_y,Q_y,P_u$ and
$Q_u$, we apply $\partial_x,\partial_y$ and $\partial_u$,
respectively, to (\ref{Pz}-\ref{Qz}) and substitute all
appearances of $R_u$ in terms of the other derivatives of interest
by the use of (\ref{Ru}). There is some freedom in choosing the
ordering of partial derivatives that converts the resulting
second-order system into a first-order one. The following results
from one of such choices:
\begin{subequations}\label{derivatives}
\begin{eqnarray}
   2\partial_uR_x
&=& \partial_xP_x+\partial_yQ_x+\partial_xR_z   ,\label{Rxu}\\
   2\partial_uR_y
&=& \partial_xP_y+\partial_yQ_y+\partial_yR_z   ,\label{Ryu}\\
   2\partial_uR_z
&=& \partial_xR_x+\partial_yR_y+\partial_zR_z   ,\label{Rzu}\\
    \partial_zP_x
&=& \partial_xR_x               ,\label{Pxz}\\
    \partial_zQ_x
&=& \partial_yR_x               ,\label{Qxz}\\
    \partial_zP_y
&=& \partial_xR_y               ,\label{Pyz}\\
    \partial_zQ_y
&=& \partial_yR_y               ,\label{Qyz}\\
   2\partial_zP_u
&=& \partial_xP_x+\partial_xQ_y+\partial_xR_z   ,\label{Puz}\\
   2\partial_zQ_u
&=& \partial_yP_x+\partial_yQ_y+\partial_yR_z   ,\label{Quz}
\end{eqnarray}
\end{subequations}

\noindent This is again a characteristic problem in
canonical form.  Notice that the first seven equations,
(\ref{Rxu}-\ref{Qyz}), decouple from the last two, since they
involve all the variables except $P_u$ or $Q_u$. Thus
$R_x,R_y,R_z,P_x,P_y,Q_x$ and $Q_y$ can be found without knowledge
of  $P_u$ or $Q_u$. Once they are known, they can be used as known
sources in the right hand side of (\ref{Puz}-\ref{Quz}) to integrate
$P_u$ and $Q_u$ via
\begin{subequations}\label{solPuQu}
\begin{eqnarray}
P_u &=& P_u|_{z=0}+\frac12\partial_x\int_0^z
    \left(P_x+Q_y+R_z\right) dz'    ,\\
Q_u &=& Q_u|_{z=0}+\frac12\partial_y\int_0^z
    \left(P_x+Q_y+R_z\right) dz'    .
\end{eqnarray}
\end{subequations}

\noindent For this reason, we now restrict attention to the
subsystem (\ref{Rxu}-\ref{Qyz}). It constitutes a characteristic
problem of the form
\begin{equation}
    \bm{B^a}\partial_aw = 0,
\end{equation}

\noindent for $w=(R_x,R_y,R_z,P_x,Q_x,P_y,Q_y)$ where the four
principal seven-dimensional matrices $\bm{B^a}$ are symmetric and
have constant coefficients, which implies a conservation law:
\begin{equation}
    \partial_a(w\bm{B^a}w) = 0.
\end{equation}

\noindent Integrating this conservation law on the ``hyper-prism'' we
have
\begin{eqnarray}
\lefteqn{\int_{u+z=T}\hspace{-0.8cm}
     w(\bm{B^u}\!+\!\bm{B^z})w \;dzdxdy =}&& \nonumber\\
&&\int_{u=0} \hspace{-0.4cm}w\bm{B^u}w \;dzdxdy
+\int_{z=0} \hspace{-0.4cm}w \bm{B^z}w \;dudxdy
\end{eqnarray}

\noindent Since $w\bm{B^u}w = 2(R_x^2+R_y^2+R_z^2)$
and $w\bm{B^z}w = -R_z^2+P_x^2+Q_x^2+P_y^2+Q_y^2$, this equation
reads explicitly
\begin{eqnarray}
&&\hspace{-1cm}\int_{\Sigma_T}
      (2R_x^2+2R_y^2+R_z^2
      +P_x^2+Q_x^2+P_y^2+Q_y^2)d\Sigma_T\nonumber\\
&=&2\int_{\Sigma_u} (R_x^2+R_y^2+R_z^2)d\Sigma_u \nonumber\\
&&
+\int_{\Sigma_z} (-R_z^2+P_x^2+Q_x^2+P_y^2+Q_y^2)d\Sigma_z ,
\end{eqnarray}

\noindent which trivially implies
\begin{eqnarray}\label{estimatederiv}
||w||^2_T
&\le& 2\Big(
   \int_{\Sigma_u} (R_x^2+R_y^2+R_z^2)d\Sigma_u \nonumber\\
&&+\int_{\Sigma_z} (P_x^2+Q_x^2+P_y^2+Q_y^2)d\Sigma_z\Big) ,
\end{eqnarray}

\noindent Here the norm of the derivatives $w$ is the natural
extension of the norm of the solution $v$ to seven dimensions,
defined by
\begin{equation}
||w||^2_T \equiv \int_{\Sigma_T}
      (R_x^2+R_y^2+R_z^2
      +P_x^2+Q_x^2+P_y^2+Q_y^2)d\Sigma_T.
\end{equation}

\noindent The inequality (\ref{estimatederiv}) gives us an
estimate of the size of the derivatives of interest in terms of
the derivatives of the free data, accounting for all of the free
data without over-counting, in a manner similar to the estimates
obtained in the case of the standard initial value problem, except
that $P_u$ and $Q_u$ are treated in a special manner, that is: as
solutions of ordinary differential equations, Eqs.
(\ref{Puz}-\ref{Quz}). Therefore, the derivatives of the solution
of the characteristic problem of the wave equation are \textit{a
priori} under control.

\section{Well-posedness of the characteristic problem of the wave
equation with a null cone data surface}
\label{sec:4}

The case of data prescribed on a null cone shares all the conceptual
features of the case of a null plane. However, it becomes technically
much more involved, essentially due to the presence of non-constant
coefficients in the expression of the wave equation in spherical
coordinates.  The estimates of the solution in terms of the data can
be derived in a manner similar as the case of a null plane, and are
obtained in Subsection~\ref{subsec:4.1}.  In order to estimate the
derivatives, however, new tools are necessary, which are developed in
Subsection~\ref{subsec:4.2}.

\subsection{Estimate of the solution}
\label{subsec:4.1}

If we wish to prescribe data on a null cone, it is convenient to
transform to spherical coordinates $(x,y,z)\to(r,\theta,\phi)$, in
which case the wave equation (\ref{waveeq}) reads
\begin{equation}
    \psi_{tt}
    -\frac{1}{r}\partial_{rr}(r\psi)
    -\frac{1}{r^2\sin\theta}\partial_\theta
    \left(\sin\theta\psi_\theta\right)
    -\frac{1}{r^2\sin^2\theta}\psi_{\phi\phi}
   =0
\end{equation}

\noindent For convenience we define the new variable
\begin{equation}
    g\equiv r\psi
\end{equation}

\noindent and change $\theta$ to a linear coordinate $\theta\to
s\equiv\cos\theta$, obtaining
\begin{equation}\label{sphericalwaveeq}
    g_{tt}
    -g_{rr}
    -\frac{1}{r^2}\partial_s
    \left((1-s^2)g_s\right)
    -\frac{1}{r^2(1-s^2)}g_{\phi\phi}
   =0.
\end{equation}

\noindent Now we change coordinates $(t,r,s,\phi)\to (u,r,s,\phi)$ via
\begin{equation}
    u=t-r,
\end{equation}

\noindent and the wave equation (\ref{sphericalwaveeq}) becomes
\begin{equation}\label{nullsphericalwaveeq}
    2g_{ur}
    -g_{rr}
    -\frac{1}{r^2}\partial_s
    \left((1-s^2)g_s\right)
    -\frac{1}{r^2(1-s^2)}g_{\phi\phi}
   =0.
\end{equation}

\noindent In order to turn it into a first-order problem we
define $R\equiv g_r$, $P\equiv g_s \sqrt{1-s^2}/r$ and  $Q\equiv
g_\phi/(\sqrt{1-s^2}r)$.  We obtain the following system
\begin{subequations}\label{nullspherical}
\begin{eqnarray}
   2 R_u &=& R_r +\frac{\sqrt{1-s^2}}{r}P_s
         +\frac{1}{r\sqrt{1-s^2}}Q_\phi \nonumber\\
&&      -\frac{sP}{r\sqrt{1-s^2}}   \label{coneRu}\\
   P_r &=& \frac{\sqrt{1-s^2}}{r}R_s -\frac{P}{r}   \label{Pr}   \\
   Q_r &=& \frac{1}{r\sqrt{1-s^2}}R_\phi-\frac{Q}{r}  \label{Qr}  \\
   g_r &=& R                \label{gr}
\end{eqnarray}
\end{subequations}

\noindent  For a unique solution to exist in the region of $R^4$
limited by a worldtube of radius $r_0$ and a null cone at $u=0$,
this system requires prescribed values of $R$ on the null surface
$\Sigma_u$ at $u=0$, and values of $g,P,Q$ on the worldtube
$\Sigma_r$ at $r=r_0$. As in the previous case, the variables
$P,Q,R$ can be obtained with no knowledge of $g$, and so can
their estimate. Ignoring (\ref{gr}), the remaining equations in
the system  (\ref{nullspherical}) have the form
\begin{equation}\label{interm}
    \bm{A^u} v,_u +\bm{A^r} v,_r+
    \bm{A^s} v,_s+
    \bm{A^\phi} v,_\phi=
    \bm{D}v
\end{equation}
where $v=(R,P,Q)$, and the matrices are given by
\begin{equation}\label{sphericalnullmatricesur}
\bm{A^u}
=
\left(\begin{array}{rrr}
    2&0&0\\
    0&0&0\\
    0&0&0
      \end{array}
\right)
\hspace{0.5cm}
\bm{A^r}
=
\left(\begin{array}{rrr}
    -1&0&0\\
    0&1&0\\
    0&0&1
      \end{array}
\right)
\end{equation}
\begin{equation}\label{sphericalnullmatricess}
\bm{A^s}
=-\frac{\sqrt{1-s^2}}{r}
\left(\begin{array}{rrr}
    0&1&0\\
    1&0&0\\
    0&0&0
      \end{array}
\right)
\end{equation}
\begin{equation}\label{sphericalnullmatricesphi}
\bm{A^\phi}
=-\frac{1}{r\sqrt{1-s^2}}
\left(\begin{array}{rrr}
    0&0&1\\
    0&0&0\\
    1&0&0
      \end{array}
\right)
\end{equation}
\begin{equation}
\bm{D}
=-\frac{1}{r}
\left(\begin{array}{ccc}
    0&s/\sqrt{1-s^2}&0\\
    0&1&0\\
    0&0&1
      \end{array}
\right)
\end{equation}

\noindent There are two difficulties with this characteristic
problem that are absent in the case of null-plane data of
Section~\ref{sec:3}. First, the principal matrices have non-constant
coefficients. This is not an obstacle, though, because, since they
are symmetric, we can still obtain a ``conservation law''. The
second difficulty is that there are undifferentiated terms, which
will appear as sources of the conservation law. Multiplying
Eq.~(\ref{interm}) by $v$ on the left, and after combining terms
appropriately, we have
\begin{equation}\label{sphericalnulldiv}
    (v\bm{A^u} v),_u + (v\bm{A^r} v),_r+
    (v\bm{A^s} v),_s+
    (v\bm{A^\phi} v),_\phi =2v\tilde{\bm{D}}v,
\end{equation}

\noindent or $\partial_aJ^a=S$ with $J^a\equiv v\bm{A^a}
v$. Here we have
\begin{equation}
\tilde{\bm{D}}
=-\frac{1}{r}
\left(\begin{array}{ccc}
    0&0&0\\
    0&1&0\\
    0&0&1
      \end{array}
\right).
\end{equation}

\noindent Since $v\tilde{\bm{D}}v=-(P^2+Q^2)/r$, it follows that
\begin{equation}
\partial_a(v\bm{A^a}v) \le 0.
\end{equation}

\begin{figure}
 \includegraphics[width=2.3in]{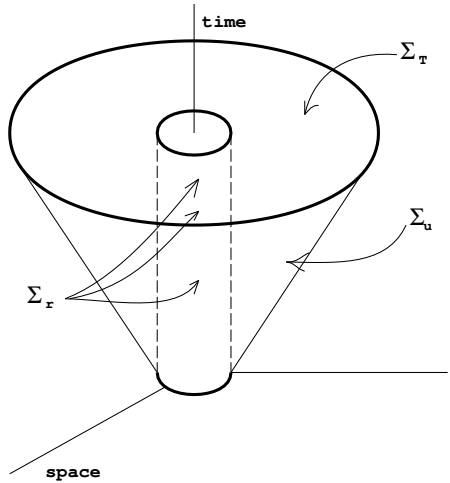}

\caption{The region of integration for  the characteristic problem
with a cone data surface. The top boundary surface $\Sigma_T$ lies
at $u+r=T+r_0$ constant and has measure $d\Sigma_T =
drdsd\phi/\sqrt{2}$.  The surface $\Sigma_u$ is a null cone at
$u=0$ and has measure $d\Sigma_u = drdsd\phi$. The surface
$\Sigma_r$ is a worldtube of radius $r=r_0$ and has measure
$d\Sigma_z = dudsd\phi$. Both surfaces $\Sigma_u$ and $\Sigma_r$
are data surfaces for the characteristic problem.  \label{fig2}}

\end{figure}

\noindent Thus the presence of the undifferentiated terms in this
particular case will not affect the estimate. We can integrate now
this law on a spacetime volume ${\cal V}^4$, and convert the volume
integral of the divergence into a surface integral over the boundary
of ${\cal V}^4$.  We choose a region of integration limited
by $\Sigma_u$ at $u=0$, $\Sigma_r$ at $r=r_0$ and $\Sigma_T$ at a
constant value of $u+r=T+r_0$, with $-1\le s\le 1$ and $0\le \phi
\le 2\pi$.  The region of integration is represented in
Figure~\ref{fig2}. We have

\begin{eqnarray}\label{intcone}
0 \ge \int_{{\cal V}^4} \partial_a(v\bm{A^a}v) dudrdsd\phi
&=&
\int_{\Sigma_T} v(\bm{A^u}+\bm{A^r})v \;drdsd\phi\nonumber\\
&&-\int_{\Sigma_u} v\bm{A^u}v \;drdsd\phi\nonumber\\
&&-\int_{\Sigma_r} v\bm{A^r}v \;dudsd\phi
\end{eqnarray}

\noindent Similarly as in the case of the null plane data surface,
we define
\begin{eqnarray}\label{defnormnullcone}
||v||_T^2 &\equiv& \int_{\Sigma_T}
    \left(R^2+P^2+Q^2\right)\;drdsd\phi\\
    &=& \int_{\Sigma_T}
    v(\bm{A^u}\!+\!\bm{A^r})v\;drdsd\phi, \nonumber
\end{eqnarray}

\noindent With (\ref{defnormnullcone}), the inequality
(\ref{intcone}) reads explicitly
\begin{eqnarray}
||v||_T^2
&\le&2\int_{\Sigma_u} R^2 \;drdsd\phi\nonumber\\
&&+\int_{\Sigma_r} \left(-R^2+P^2+Q^2 \right)\;dudsd\phi
\end{eqnarray}

\noindent which by the same argument as in the case of the null
plane data surface leads to a similar kind of estimate
\begin{equation}\label{estimatenullcone}
 ||v||_T^2 \le
    2\left(\!\int_{\Sigma_u}\hspace{-0.4cm}R^2 d\Sigma_u
       +\int_{\Sigma_r}\hspace{-0.35cm}
        \left(P^2+Q^2\right)
    d\Sigma_r\right).
\end{equation}

\noindent So we see that the ``size'' of the data on both data
surfaces controls the ``size'' of the solution. Finally, once $R$
is known, the variable $g$ can be found and estimated by
integrating (\ref{gr}) with given values on $r=r_0$:
\begin{equation}
    g = g|_{r_0} + \int_{r_0}^r R dr'.
\end{equation}

\subsection{Estimates of the derivatives}
\label{subsec:4.2}

The results of the previous Subsection establish that the size of
the solution is controlled by the size of the free data. We wish
to be able to control the size of the derivatives $R_r,R_s,R_\phi,
P_u,P_s,P_\phi, Q_u,Q_s$ and $Q_\phi$ of the solution in terms of
the derivatives of the free data, as well, since these are the
derivatives that play a role analogous to that of the space
derivatives in the case of the Cauchy problem. This will be
possible if we can write down a characteristic system of equations
for these nine variables, where $R_r,R_s,R_\phi$ will be normal
variables and $P_u,P_s,P_\phi, Q_u,Q_s$ and $Q_\phi$ will be null
variables, as a consequence of the original system of equations.
If we write down a system for the derivatives themselves in a
manner completely analogous to the method employed in
Subsection~\ref{subsec:3.2}, the system (\ref{nullspherical})
implies the following nine equations:
\begin{widetext}
\begin{subequations}\label{prederivativesnullcone}
\begin{eqnarray}
   2\partial_uR_s
- \frac{\sqrt{1-s^2}}{r}\partial_sP_s
- \frac{1}{r\sqrt{1-s^2}}\partial_\phi Q_s
- \partial_s R_r
&=&
-\frac{2sP_s}{r\sqrt{1-s^2}}
+\frac{sQ_\phi-(1+s^2)P}{r(1-s^2)^\frac32}  ,\label{Rsu}\\
   2\partial_u R_\phi
- \frac{\sqrt{1-s^2}}{r}\partial_sP_\phi
- \frac{1}{r\sqrt{1-s^2}}\partial_\phi Q_\phi
- \partial_\phi R_r
&=&
-\frac{sP_\phi}{r\sqrt{1-s^2}}      ,\label{Rphiu}\\
   2\partial_uR_r
- \frac{1-s^2}{r^2}\partial_sR_s
- \frac{1}{r^2(1-s^2)}\partial_\phi R_\phi
- \partial_s R_r
&=&
\frac{sR_s}{r}
-\frac{2\sqrt{1-s^2}P_s}{r^2}
+\frac{s(P-rP_r)-2Q_\phi}{r^2\sqrt{1-s^2}}  ,\label{Rru}\\
    \partial_rP_s
- \frac{\sqrt{1-s^2}}{r}\partial_sR_s
&=&
-\frac{P_s}{r}-\frac{sR_s}{r\sqrt{1-s^2}}   ,\label{Psr}\\
    \partial_rQ_s
- \frac{1}{r\sqrt{1-s^2}}\partial_\phi R_s
&=&
-\frac{Q_s}{r}-\frac{sR_\phi}{r(1-s^2)^{\frac32}},\label{Qsr}\\
    \partial_rP_\phi
- \frac{\sqrt{1-s^2}}{r}\partial_sR_\phi
&=&
-\frac{P_\phi}{r}               ,\label{Pphir}\\
    \partial_rQ_\phi
- \frac{1}{r\sqrt{1-s^2}}\partial_\phi R_\phi
&=&
-\frac{Q_\phi}{r}               ,\label{Qphir}\\
   2\partial_rP_u
-\frac{\sqrt{1-s^2}}{r}\partial_sR_r
-\frac{1-s^2}{r^2}\partial_sP_s
-\frac{1}{r^2}\partial_sQ_\phi
&=&
-\frac{2sP_s}{r^2}
+\frac{sQ_\phi-(1+s^2)P}{r^2(1-s^2)}
-\frac{2P_u}{r}                 ,\label{Pur}\\
   2\partial_rQ_u
-\frac{1}{r\sqrt{1-s^2}}\partial_\phi R_r
-\frac{1}{r^2}\partial_\phi P_s
-\frac{1}{r^2(1-s^2)}\partial_\phi Q_\phi
&=&
-\frac{sP_\phi}{r^2(1-s^2)}
-\frac{2Q_u}{r}                 ,\label{Qur}
\end{eqnarray}
\end{subequations}
\end{widetext}

\noindent The first seven equations do not involve $P_u$ nor
$Q_u$; therefore, we may consider Eqs.~(\ref{Rsu})-(\ref{Qphir})
as a system of seven equations for the seven variables
$R_r,R_s,R_\phi,P_s,P_\phi,Q_s$ and $Q_\phi$, the solution of
which can be used as a known source for (\ref{Pur})-(\ref{Qur}).
For the moment, we ignore the last two equations, which we will
come back to once we have an estimate from the first seven. The
system (\ref{Rsu})-(\ref{Qphir}) has many features that make it
unsuitable for our purposes. In the first place, the
undifferentiated function $P$ is involved, but it is not one of
our nine variables. This can be interpreted as a non-homogeneous
system of equations for the seven first derivatives of interest,
where $P$ acts as a known forcing source. $P$'s presence
would definitely affect any estimate that may be implied by the
system as it stands.  Second, there are many undifferentiated
terms in these equations, making it almost certain that they will
play a role in any estimates. Thirdly, the principal matrices are
not symmetric, although their asymmetry is not severe because the
vanishing coefficients appear symmetrically. As we know, the
symmetry of the principal matrices would lead directly to a
conservation law with a non-vanishing source of undifferentiated
variables, our main goal.

The first obstacle is a true impediment as far as we can see. The
standard way to derive estimates for non-homogeneous systems of
equations is by means of Duhamel's principle, which allows one to
express the solution of the non-homogeneous system in terms of the
solution of the associated homogeneous system and forcing source
function~\cite{kreissbook}. We are not aware of any analogous
principle for the characteristic problem. Thus, at the moment we
are forced to consider only homogeneous characteristic problems.
Therefore, we must find a choice of fundamental variables for this
system which removes the non-homogeneous terms.

Fortunately, there exists a choice of fundamental variables for
Eqs.~(\ref{Rsu})-(\ref{Qphir}) which at the same time symmetrizes
the principal matrices and eliminates the appearance of the
undifferentiated function $P$, thus taking care of the two most
important deficiencies of the system of evolution for the
derivatives. Instead of using the seven first derivatives as
fundamental variables, we can use the following:
\begin{eqnarray}
\widehat{R}^\phi \equiv \frac{R_\phi}{r\sqrt{1-s^2}}, \qquad
\widehat{R}^s &\equiv &\frac{R_s\sqrt{1-s^2}}{r} ,\\
\widehat{P}^\phi \equiv \frac{P_\phi}{r\sqrt{1-s^2}}, \qquad
\widehat{P}^s &\equiv
&\frac{\partial_s\!\!\left(\!P\sqrt{1-s^2}\right)}{r}\\
\widehat{Q}^\phi \equiv \frac{Q_\phi}{r\sqrt{1-s^2}},\qquad
\widehat{Q}^s &\equiv &\frac{Q_s\sqrt{1-s^2}}{r}.
\end{eqnarray}

\noindent This re-scaling will necessarily contribute more
undifferentiated terms to the system because of the dependence of
the coefficients on $r$ and $s$. Additionally, the
undifferentiated function $P$ is absorbed into the fundamental
variable $\widehat{P}^s$. In terms of the variables $w = (w_i)
\equiv (\widehat{R}^s,
\widehat{R}^\phi,R_r,\widehat{P}^s,\widehat{P}^\phi,
\widehat{Q}^s,\widehat{Q}^\phi)$, the system
(\ref{Rsu})-(\ref{Qphir})has the form
\begin{equation}\label{preconsderivnullcone}
\bm{B^a}\partial_a w = \bm{D}w,
\end{equation}

\noindent where all the principal matrices $\bm{B^a}$ are symmetric
and given explicitly by $B^u{}_{ij}= diagonal(2,2,2,0,0,0,0)$,
$B^r{}_{ij}= diagonal(0,0,-1,1,1,1,1)$ and $B^s{}_{ij} =
B^\phi{}_{ij} =0$ except
\begin{eqnarray}
B^s{}_{13}&=&B^s{}_{14}=B^s{}_{26}= -\frac{\sqrt{1-s^2}}{r}\\
B^\phi{}_{15}&=&B^s{}_{23}=B^s{}_{27}= -\frac{1}{r\sqrt{1-s^2}}
\end{eqnarray}

\noindent The matrix $\bm{D}$ has coefficients that depend on $r$
and $s$ but not on the unknown variables. The explicit expressions
of $D_{ij}$ are of no relevance for our purposes because
they are sufficiently generic to force us to consider the most
general case. Multiplying by $w$ on the left,
Eq.~(\ref{preconsderivnullcone}) implies
\begin{equation}\label{next}
\partial_a(w\bm{B^a}w) = w\tilde{\bm{D}}w,
\end{equation}

\noindent with $\tilde{\bm{D}}=2 \bm{D} + \partial_a\bm{B^a}$.
Integrating (\ref{next}) on the volume ${\cal V}^4$ enclosed by
$\Sigma_T, \Sigma_u$ and $\Sigma_r$ we find
\begin{eqnarray}\label{next2}
&&\int_{\Sigma_T}\hspace{-0.4cm}w(\bm{B^u}\!+\!\bm{B^r})w \,d\Sigma_T
-\int_{\Sigma_u}\hspace{-0.4cm}w\bm{B^u}w\,d\Sigma_u
-\int_{\Sigma_r}\hspace{-0.4cm}w\bm{B^r}w\,d\Sigma_r\nonumber\\
&&=
\int_{{\cal V}^4}w\tilde{\bm{D}}w\, d{\cal V}^4
\end{eqnarray}

\noindent We define
\begin{equation}
||w||^2_T = \int_{\Sigma_T} w^2 \,d\Sigma_T
\end{equation}

\noindent with $w^2\equiv\sum_iw_i^2$. Then, exactly as in
Subsection~\ref{subsec:3.2}, and because $\bm{B^u}\!+\!\bm{B^r}$ is
the identity matrix, Eq.~(\ref{next2}) implies
\begin{eqnarray}\label{next3}
||w||_T &\le& 2 \Big(
\int_{\Sigma_u}\hspace{-0.4cm}
    R_r^2\!+\!\widehat{R}^s{}^2
        \!+\!\widehat{R}^\phi{}^2d\Sigma_u\nonumber\\
&&
+\int_{\Sigma_r}\hspace{-0.4cm}
    \widehat{P}^s{}^2
    \!+\!\widehat{P}^\phi{}^2
    \!+\!\widehat{Q}^s{}^2
    \!+\!\widehat{Q}^\phi{}^2d\Sigma_r\Big)  \nonumber\\
&&+ \int_{{\cal V}^4}w\tilde{\bm{D}}w\,d{\cal V}^4
\end{eqnarray}

\noindent The presence of the undifferentiated terms is a
complication to the estimate of $w$ in terms of the free data on
both surfaces. This complication, however, can be resolved in a
manner analogous as the case of the initial value problem dealt
with in Section~\ref{sec:2}. First notice that
\begin{equation}
w\tilde{\bm{D}}w \le c w^2
\end{equation}

\noindent where $c=\mbox{max}(|\tilde{D}_{ij}|)$ in the volume
${\cal V}^4$. Thus $\int_{{\cal V}^4} w\tilde{\bm{D}}w d{\cal
V}^4\le c\int_0^Tdt \int_{\Sigma_t}w^2d\Sigma_t$. With this, the
inequality (\ref{next3}) implies
\begin{eqnarray}\label{next4}
||w||^2_T &\le& 2 \Big( \int_{\Sigma_u}\hspace{-0.4cm}
    R_r^2\!+\!\widehat{R}^s{}^2
        \!+\!\widehat{R}^\phi{}^2d\Sigma_u\nonumber\\
&&
+\int_{\Sigma_r}\hspace{-0.4cm}
    \widehat{P}^s{}^2
    \!+\!\widehat{P}^\phi{}^2
    \!+\!\widehat{Q}^s{}^2
    \!+\!\widehat{Q}^\phi{}^2d\Sigma_r\Big)  \nonumber\\
&&+ c\int_0^T ||w||^2_t dt
\end{eqnarray}

\noindent This holds for any fixed value of $T$, where $\Sigma_u$
and $\Sigma_r$ extend as far as their intersection with the
surface at fixed value of $u+r=T$. If we denote $\Sigma_{u,t}$ and
$\Sigma_{r,t}$ the regions of $\Sigma_u$ and $\Sigma_r$ extending
only so far as their intersection with $u+r=t\le T$, we can write
\begin{eqnarray}\label{next5}
||w||^2_t &\le& 2 \Big( \int_{\Sigma_{u,t}}\hspace{-0.4cm}
    R_r^2\!+\!\widehat{R}^s{}^2
        \!+\!\widehat{R}^\phi{}^2d\Sigma_{u,t}\nonumber\\
&&
+\int_{\Sigma_{r,t}}\hspace{-0.4cm}
    \widehat{P}^s{}^2
    \!+\!\widehat{P}^\phi{}^2
    \!+\!\widehat{Q}^s{}^2
    \!+\!\widehat{Q}^\phi{}^2d\Sigma_{r,t}\Big)  \nonumber\\
&&+ c\int_0^t ||w||^2_{t'} dt' \nonumber\\
&\le&2 \Big( \int_{\Sigma_u}\hspace{-0.4cm}
    R_r^2\!+\!\widehat{R}^s{}^2
        \!+\!\widehat{R}^\phi{}^2d\Sigma_u\nonumber\\
&&
+\int_{\Sigma_r}\hspace{-0.4cm}
    \widehat{P}^s{}^2
    \!+\!\widehat{P}^\phi{}^2
    \!+\!\widehat{Q}^s{}^2
    \!+\!\widehat{Q}^\phi{}^2d\Sigma_r\Big)  \nonumber\\
&&+ c\int_0^t ||w||^2_{t'} dt',
\end{eqnarray}

\noindent which holds for any value of $t\le T$.
Using this inequality recursively into the right-hand side of
(\ref{next4}) we have
\begin{eqnarray}\label{next6}
||w||^2_T &\le& 2
\left(1+cT+\frac{c^2T^2}{2}+...+\frac{c^mT^m}{m!}\right)\nonumber\\
&&\times\Big(
 \int_{\Sigma_u}\hspace{-0.4cm}
    R_r^2\!+\!\widehat{R}^s{}^2
        \!+\!\widehat{R}^\phi{}^2d\Sigma_u\nonumber\\
&&
+\int_{\Sigma_r}\hspace{-0.4cm}
    \widehat{P}^s{}^2
    \!+\!\widehat{P}^\phi{}^2
    \!+\!\widehat{Q}^s{}^2
    \!+\!\widehat{Q}^\phi{}^2d\Sigma_r\Big)  \nonumber\\
&&+ c^{m+1}\int_0^T\!\!\!dt_1\int_0^{t_1}\!\!\!dt_2...\int_0^{t_m}
\hspace{-0.4cm}||w||^2_{t_{m+1}} dt_{m+1}
\end{eqnarray}

\noindent for any given non-negative integer $m$. In the limit for
$m\to\infty$ the sequence in the right-hand side converges if
$cT<1$, in which case we have
\begin{eqnarray}\label{next7}
||w||^2_T &\le& 2 e^{cT}\Big(
 \int_{\Sigma_u}\hspace{-0.4cm}
    R_r^2\!+\!\widehat{R}^s{}^2
        \!+\!\widehat{R}^\phi{}^2d\Sigma_u\nonumber\\
&& +\int_{\Sigma_r}\hspace{-0.4cm}
    \widehat{P}^s{}^2
    \!+\!\widehat{P}^\phi{}^2
    \!+\!\widehat{Q}^s{}^2
    \!+\!\widehat{Q}^\phi{}^2d\Sigma_r\Big)
\end{eqnarray}

\noindent This is the final estimate for the derivatives in terms
of the derivatives of the free data on both surfaces. Like in the
case of the initial value problem dealt with in
Section~\ref{sec:2}, the estimate involves an exponential factor,
essentially due to the presence of undifferentiated terms in the
system of equations for $w$. As usual in such cases, even though
the estimate is useful in order to prove that the solution depends
continuously on the data for any value of $T$, it is impractical
for large $T$ for the purpose of estimating the error in a
numerical solution. In particular, our proof only guarantees the
estimate for $T<c^{-1}$. Perhaps with greater care, possibly by
using Gronwall's inequality in integral form, the estimate could
be extended to longer values of $T$.

Since the seven derivatives of interest can be found and estimated
independently of $P_u$ and $Q_u$, they can now be used as known
sources for Eqs.~(\ref{Pur})-(\ref{Qur}) written in terms of the new
fundamental variables. The solutions $P_u$ and
$Q_u$ can be obtained by quadratures as
\begin{eqnarray}
P_u &=& \frac{r_0}{r}P_u|_{r_0}
       +\frac{\sqrt{1-s^2}}{2r}\partial_s\!\!\int_{r_0}^r
\!\!\!
R_r\!+\!\widehat{P}^s\!\!+\!\widehat{Q}^\phi dr',\\
Q_u &=& \frac{r_0}{r}Q_u|_{r_0}
       +\frac{1}{2r\sqrt{1-s^2}}\partial_\phi\!\!\int_{r_0}^r
\!\!\!
R_r\!+\!\widehat{P}^s\!\!+\!\widehat{Q}^\phi dr',
\end{eqnarray}

\noindent from the known functions and from free data given on
$r=r_0$.

\section{Concluding remarks}
\label{sec:5}

Summarizing, Sections~\ref{sec:3} and ~\ref{sec:4} develop the
proofs of the following two theorems:

\begin{theorem}

Consider Eqs.~(\ref{Ru})-(\ref{Qz}), representing the first-order
reduction of the wave equation in three spatial cartesian
coordinates $(x,y,z)$ and one null coordinate $u=t-z$. Given data
$R|_{u=0} = f(x,y,z)$ on the null surface $\Sigma_u$ at $u=0$
with$0\le z\le T$, $P|_{z=0}=g(u,x,y)$ and $Q|_{z=0}=h(u,x,y)$ on
the timelike surface $\Sigma_z$ at $z=0$ with $0\le u\le T$, the
unique solution $v=(R,P,Q)$ periodic in $(x,y)$ satisfies the
estimate
\[
 ||v||_T^2 \le 2\left(\!\int_{\Sigma_u}\hspace{-0.4cm}R^2 d\Sigma_u
       +\int_{\Sigma_z}\hspace{-0.35cm}
        \left(P^2+Q^2\right)
    d\Sigma_z\right).
\]

\noindent with $||v||_T^2 \equiv
\int_{\Sigma_T}(R^2+P^2+Q^2)d\Sigma_T$, where $\Sigma_T$ is the
spacelike surface $u+z=T$ for $0\le u\le T$ and $0\le z\le T$.

The derivatives of $v$ are similarly bounded by the derivatives of
$f,g$ and $h$.

\end{theorem}

\begin{theorem}

Consider Eqs.~(\ref{coneRu})-(\ref{Qr}), representing the
first-order reduction of the wave equation in three spatial
spherical coordinates $(r,s=\cos\theta,\phi)$ and one null
coordinate $u=t-r$. Given data $R|_{u=0} = f(r,s,\phi)$ on the null
surface $\Sigma_u$ at $u=0$ with $r_0\le r\le T+r_0$,
$P|_{r=r_0}=g(u,s,\phi)$ and $Q|_{r=r_0}=h(u,s,\phi)$ on the
timelike surface $\Sigma_r$ at $r=r_0$ with $0\le u\le T$, the
unique solution $v=(R,P,Q)$ satisfies the estimate
\[
 ||v||_T^2 \le 2\left(\!\int_{\Sigma_u}\hspace{-0.4cm}R^2 d\Sigma_u
       +\int_{\Sigma_z}\hspace{-0.35cm}
        \left(P^2+Q^2\right)
    d\Sigma_r\right).
\]

\noindent with $||v||_T^2 \equiv
\int_{\Sigma_T}(R^2+P^2+Q^2)d\Sigma_T$, where $\Sigma_T$ is the
spacelike surface $u+r=T+r_0$ for $0\le u\le T$ and $r_0\le r\le
T+r_0$.

The derivatives of $v$ are similarly bounded by the derivatives of
$f,g$ and $h$ for small values of $T$.

\end{theorem}

These results are relevant to the stability of the solutions of the
wave equation constructed from data given on two intersecting
transverse surfaces, one of which is timelike and the other one is
characteristic. The theorems guarantee that the solutions will be
stable under small perturbations of the data on such two surfaces.
Since the existence and uniqueness of solutions is already guaranteed
by Duff's theorem~\cite{duff},  our result generalizes the standard
notion of well-posedness, available in the context of Cauchy problems,
to the characteristic problem of the wave equation.

In the wider context of general interest in characteristic problems of
any kind, three publications  posterior to Duff's pioneering
work~\cite{duff} stand out for relevance and motivation. In the first
place,M\"{u}ller zum Hagen \& Seifert~\cite{zumhagen} recognized the
value of energy estimates and correctly characterize the data surfaces
and their role in the estimates for all types of problems, including
problems with one or more characteristic surfaces. Perhaps due to the
generic nature of their work, their estimates appear to have treated
indifferently the normal data and the null data, making no distinction
between free data and data that propagates within each characteristic
surfaces. This problem is pointed out by Rendall~\cite{rendall}, who,
in reference to the M\"{u}ller zum Hagen \& Seifert work,  says that
``they attempted, not entirely successfully, to give an existence and
uniqueness proof by following step by step the treatment of the Cauchy
problem.''  Rendall proceeds to detail arguments for existence,
uniqueness and stability for symmetric hyperbolic systems --like the
wave equation-- in the case of two characteristic transverse data
surfaces. Rendall does conclude that estimates exist on the basis of
the well posedness of the associated Cauchy problem, but
unfortunately, offers no explicit estimates of the solution in terms
of the set of free data, which he identifies.
Balean~\cite{baleanthesis,balean1,balean2} later calculates estimates
of the energy kind for the wave equation with one characteristic and
one timelike data surface where the null data are treated as a flow of
information across a timelike boundary, and the distinction between
the contributions of the free data and the transported data is not
markedly emphasized. All three works deal exclusively with
second-order equations, leaving Duff's systematic approach to
first-order characteristic problems without a follow-up --in a
strictly formal sense. In fact, we have been unable to identify any
other published literature dealing with estimates for characteristic
problems of any kind. Sometimes the second-order formulation of a
problem and its associated first-order reduction are regarded as
equally valid and interchangeable, but we find that new insights are
often to be gained when a problem (even a well-understood one) is
viewed from a different vantage point, a motivation that underlies our
current work.

The method used here to address the question of stability is quite
sensitive to the presence of undifferentiated terms and is likely
to be sensitive to the presence of non-linear terms as well.
Still, its strength lies in its conceptual features, which depart
quite significantly from the three predecessors already referred
to. It is to be hoped that our conceptual framework will be useful
as a general guideline for other characteristic problems. Work is
in progress~\cite{lincharac} to extend the method to generic
linear characteristic problems for first-order systems of
equations in order to obtain a criterion for ``manifest
well-posedness'' that would play a role analogous to that of
symmetric hyperbolicity of Cauchy problems. Our ultimate goal is
to develop new insights into the nature of the characteristic
problem of the Einstein equations in the form pioneered by
Sachs~\cite{sachs}. In this respect, see
\cite{lui,qlincharac,friedlander,helmut81a,helmut81b,helmut82} and
also \cite{jefflr} for a review including the numerical
implementation.

\begin{acknowledgments}

I am deeply indebted to Roberto G\'{o}mez for many enlightening
conversations. This work was supported by NSF under grants No.
PHY-9803301 and No. PHY-0070624 to Duquesne University.

\end{acknowledgments}


\end{document}